\documentclass[twocolumn]{aastex701}

\usepackage{amsmath}

\shorttitle{F-corona observed by Artemis II}
\shortauthors{Tsumura \& Arimatsu}
\submitjournal{ApJL}

\begin{document}

\title{Large-scale Morphology of the Optical F-corona from a Total Solar Eclipse Observation During the Artemis II Lunar Flyby}

\author[orcid=0000-0001-7143-6520]{Kohji Tsumura}
\affiliation{Department of Natural Sciences, Faculty of Science and Engineering, Tokyo City University, \\
1-28-1, Tamazutsumi, Setagaya, Tokyo 158-8557, Japan}
\affiliation{Frontier Research Institute for Interdisciplinary Sciences, Tohoku University, \\
6-3, Aramaki aza Aoba, Aoba, Sendai, Miyagi 980-8578, Japan}
\email[show]{tsumura@astr.tohoku.ac.jp}  

\author[orcid=0000-0003-1260-9502]{Ko Arimatsu} 
\affiliation{National Astronomical Observatory of Japan, 2-21-1 Osawa, Mitaka, Tokyo 181-8588, Japan}
\affiliation{Ishigakijima Astronomical Observatory, 1024-1 Arakawa, Ishigaki, Okinawa 907-0024, Japan}
\email{ko.arimatsu@nao.ac.jp}

\begin{abstract}
We investigated the structure of the optical F-corona, i.e., inner zodiacal light, using a publicly released wide-field image of a total solar eclipse that was obtained during the Artemis~II crewed lunar flyby.
In this image, the solar disk is fully occulted by the Moon, providing a rare view of diffuse circumsolar emission over a wide angular extent. Although the dataset is derived from a rendered RGB JPEG image without full photometric calibration, the gamma correction inherent to the image format was explicitly accounted for and the instrumental response was validated using field stars. The stellar calibration demonstrates a linear response within the unsaturated regime relevant to our measurements, enabling a reliable analysis of the relative morphology and brightness profiles of the F-corona.
The observed F-corona exhibits a flattened, nearly elliptical morphology aligned with the ecliptic plane, with flattening indices of 0.52, 0.54, and 0.56 for the red, green, and blue channels, respectively. Radial intensity profiles along ecliptic longitude and latitude are well described by power laws in solar elongation, although the derived slopes are systematically steeper than previous observations.
Comparison with the \texttt{ZodiSURF} zodiacal light model indicates that the observed radial profile of the F-corona along the ecliptic longitude is modestly reproduced by the model, supporting a radial dust number-density power-law index of $\alpha \sim 1.3$, even in regions near the Sun.
In a broader historical context, these results provide an empirical proof-of-concept that supports future solar coronal occultation observations from lunar orbit.
\end{abstract}


\section{Introduction} \label{sec:intro}
Zodiacal light (ZL) is a major contributor to the diffuse brightness of the night sky arising from astrophysical sources outside the terrestrial atmosphere \citep{Leinert1998, Mattila2019}. 
It results from the scattering of optical and near-infrared photospheric light from the Sun by interplanetary dust particles (IDPs); therefore, ZL observations are important for investigating the physical properties and spatial distributions of IDPs in the solar system. 

Space-based observations are required to study ZL since strong terrestrial atmospheric emissions obscure the weaker ZL signals \citep{Gary1973, Murdock1985, Matsuura1995, Matsumoto1996, Tsumura2010, Tsumura2013, BUFFINGTON201688, Korngut2022, Takimoto2022, Takimoto2023}.
A detailed model of the IDP distribution has been established based on all-sky infrared imaging by the Diffuse Infrared Background Experiment (DIRBE) onboard the Cosmic Background Explorer (COBE) \citep{Kelsall1998}.
The \texttt{ZodiSURF} model estimates the ZL brightness at optical wavelengths by combining the DIRBE-based IDP distribution with a scattering phase function and albedo empirically determined from the sky surface brightness measured by the Hubble Space Telescope (HST) \citep{OBrien_2026}.
However, the applicability of this model is limited to solar elongation angles greater than $80^\circ$.

The Fraunhofer corona (F-corona; \citealt{Hulst1947}) is understood as the inner
extension of the ZL rather than as a distinct phenomenon.
Observing the F-corona is challenging due to the strong contrast with the bright solar photosphere.
Therefore, observational opportunities are largely limited to solar eclipses or eclipse-like occultation geometries.
Space-based observations of the F-corona are reviewed by \citet{Lamy2022}. 

Historically, observations of the outer corona were performed from the lunar surface shortly after local sunset or before sunrise by four out of the seven unmanned Surveyor Moon probes between 1966 and 1968 \citep{Norton1967, Bohlin1971}, demonstrating the advantage of using the lunar disk as a natural occulter to suppress stray light.
Recently, Clementine mapped the inner ZL using lunar occultation at solar elongation angles of approximately $5^\circ$--$30^\circ$ \citep{Hahn2002}. 
The Heliospheric Imagers 1A (HI-1A) onboard the Solar Terrestrial Relations Observatory Ahead (STEREO-A) subsequently characterized the morphology and radial gradients of the inner dust cloud over solar elongation angles of $5^\circ$--$24^\circ$ \citep{Stenborg2018}, while the Large Angle Spectroscopic Coronagraph (LASCO) onboard the Solar and Heliospheric Observatory (SOHO) established a photometrically calibrated reference view of the F-corona from 2 to 30~$R_\odot$ \citep{Lamy2022}.
Additionally, the Wide-field Imager for Solar Probe (WISPR) onboard the Parker Solar Probe (PSP) extended observational coverage and found evidence for dust depletion in the innermost region of the dust cloud \citep{Stenborg2021}. Collectively, these results show that the inner dust cloud generally exhibits a flattened morphology that is approximately aligned with the symmetry plane of the zodiacal cloud; however, these observational constraints are limited due to the reliance on specialized coronagraphs or infrequent eclipse-like observational geometries.

Therefore, the use of the Moon as a stable, low-stray-light platform for systematic coronal observations has been revisited in the context of future mission concepts. \citet{Habbal2013} described the modern concept of performing regular solar coronal occultation observations from lunar orbit, emphasizing on the continuous access to coronal heights that remain difficult to observe with conventional coronagraphs. Building on this idea, \citet{Cooper2023} proposed a dedicated lunar-orbit mission architecture optimized for white-light coronal observations, explicitly designed to minimize stray light and enable repeatable measurements of the F-corona.

In this Letter, we examine the large-scale morphology of the F-corona at optical
wavelengths using a publicly available wide-field eclipse image obtained during the Artemis~II lunar flyby.
The present study provides an empirical demonstration of the scientific viability of lunar-based occultation concepts and can be regarded as a valuable proof-of-concept for future missions with a similar background.

\section{Data Acquisition} \label{sec:acquisition}
\subsection{Artemis II} \label{sec:artemis}

Artemis~II was the first crewed flight of NASA's Artemis program  \citep{NASA2026_Press}. The mission launched four astronauts aboard the Orion spacecraft on a Space Launch System (SLS) rocket from the Kennedy Space Center on April 1, 2026, at 22:35~UTC, executing an approximately 10-day free-return trajectory around the Moon and back to Earth \citep{nasa_launch_artemis2_2026, nasa_artemis2_mission_2026}. 
On flight day~6, Orion passed behind the Moon and reached its closest approach, i.e., around 6545~km from the lunar surface, at approximately 23:00 (in the
live flight log) to 23:02 (in the pre-flyby Lunar Targeting Plan) UTC.
It then entered nearly hour-long solar eclipse interval beginning at 00:35~UTC \citep{nasa_flightday6_ready_2026, nasa_flightday6_updates_2026}. 
The eclipse image under consideration was obtained after the closest approach to the Moon while Orion was returning toward Earth.

\subsection{Observation Conditions} \label{sec:condition}
\begin{figure*}[ht!]
\plotone{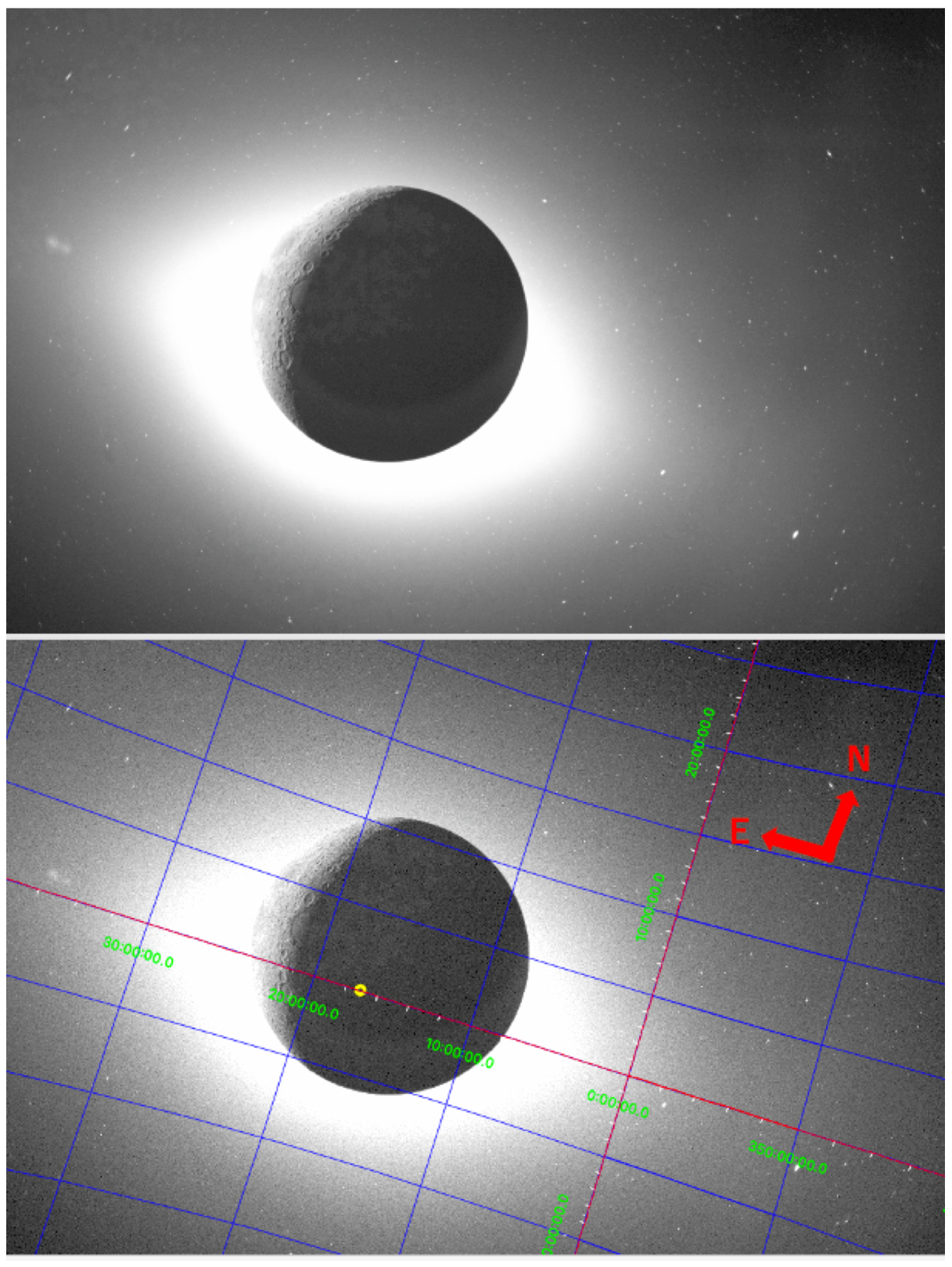}
\caption{
Top: Image \texttt{art002e009301} captured by the Artemis~II crew during their lunar flyby at an altitude of approximately 6545~km above the lunar surface.
The Moon fully occults the Sun, and the left side of the lunar disk is illuminated by sunlight reflected from the Earth.
The apparent lunar diameter is $16.9^{\circ}$.
The glowing halo around the dark lunar disk corresponds to the F-corona (inner zodiacal light), and numerous stars are visible in the surrounding field.
Bottom: The same image with an overlaid ecliptic coordinate grid.
The yellow marker indicates the position of the Sun, which has an apparent diameter of $31.9\arcmin$ and is located behind the Moon.
}
\label{original}
\end{figure*}

In this study, we analyze the NASA image \texttt{art002e009301} (Figure~\ref{original} [top]) from the Artemis~II lunar-flyby image set. 
The image was publicly released by NASA on April~7, 2026; the accompanying public caption attributes it to the Artemis~II crew during the April~6, 2026, lunar flyby. 
According to NASA's official description, the image presents a view of a total solar eclipse obtained during the lunar flyby, showing a luminous halo surrounding the dark lunar disk.

Publicly available metadata associated with the released image indicate that it was captured using a \texttt{NIKON Z9} camera with a \texttt{35~mm f/2D} lens at an aperture of $f/2$, an exposure time of $t_{\rm exp}=2$~s, and ISO~1600. 
The acquisition time was 01:06:19~UTC on April~7, 2026, i.e., approximately 2.1~h  after Orion's closest approach to the Moon, placing the exposure at
approximately 31~min after the start of the eclipse phase.

The image characteristics and observing geometry adopted in this work are summarized in Table~\ref{tab:artemis_image_context}.

\begin{deluxetable*}{ll}
\tablecaption{Observation descriptors for the analyzed Artemis~II image \label{tab:artemis_image_context}}
\tablehead{
\colhead{Descriptor} & \colhead{Value}
}
\startdata
Image identifier & \texttt{art002e009301} \\
Mission & Artemis~II (crewed Orion lunar flyby on a free-return trajectory) \\
Mission phase & Flight day~6, solar-eclipse segment of the lunar flyby \\
Image acquisition time & 2026-04-07 01:06:19~UTC \\
Public image format & RGB JPEG, $8256\times5504$ pixels, 8-bit per channel \\
Camera metadata & NIKON Z~9, 35~mm \texttt{f/2D}, $f/2$, $2$~s, ISO~1600 \\
Frame geometry & FoV $\sim51^{\circ}\times33^{\circ}$; plate scale $20\arcsec$-- $25\arcsec$~pixel$^{-1}$ \\
Solar/lunar apparent diameters & $31.9\arcmin$ / $16.9^{\circ}$ \\
Solar position from Orion & $(\lambda,\beta)\approx(17.0^{\circ},0.013^{\circ})$ in ecliptic coordinates \\
\enddata
\tablecomments{Mission phase is based on NASA Artemis~II flight updates and the Lunar Targeting Plan. Camera and file-format information were taken from the metadata of the original public JPEG.}
\end{deluxetable*}

\ \\
\ \\
\ \\

\section{Data Reduction} \label{sec:reduction}

\subsection{Image Preparation and Astrometric Registration}
\label{subsec:wcs}

Image \texttt{art002e009301} was converted into a three-plane FITS format while preserving the original pixel sampling of the RGB product. 
Since the Moon completely occults the solar disk in this observation, the effective inner boundary for the diffuse emission analysis is defined by the apparent lunar limb rather than by an instrumental occulter. 
The large apparent lunar diameter of $16.9^{\circ}$, which was calculated using \texttt{JPL Horizons} \citep{Giorgini1996}, naturally suppresses direct photospheric emission over a substantial angular range around the Sun.

The stellar field outside the lunar limb provides the astrometric reference points required to resolve the wide-field image geometry. 
An astrometric solution was derived using \texttt{Astrometry.net} \citep{Lang2010}, which uses detected field stars to recover a world coordinate system (WCS) for the image. 
The result is displayed in Figure~\ref{original} (bottom) and yields an effective field of view (FoV) of approximately $51^{\circ}\times33^{\circ}$, consistent with the values summarized in Table~\ref{tab:artemis_image_context}. 
The pixel scale varies from $20\arcsec$ at the edge of the FoV to $25\arcsec$ near its center due to optical distortion in the camera system.

The image analyzed in this work is a processed JPEG product generated by a consumer camera. 
Such images typically undergo standard in-camera corrections for instrumental effects such as large-scale vignetting. 
However, JPEG images have previously been used in astronomical image-based morphological or structural analyses when absolute radiometric calibration is not required, particularly in citizen-science projects \citep{Lintott2008, Krista2015}.
Furthermore, residual large-scale sensitivity variations that may not be fully removed by standard processing are expected to be absorbed within the empirical calibration using field stars described in Section \ref{subsec:starcal}.

\subsection{Zero-point Correction}
\label{subsec:baseline}

The released JPEG image is an 8-bit, rendered RGB product generated by a consumer camera.
Therefore, it lacks the detector-level information required for a rigorous end-to-end photometric calibration. 
Particularly, the publicly accessible file does not preserve the raw camera response, and the bias, dark-current, flat-field, compression, white-balance, and color-rendering steps applied in the public processing chain are not fully documented. 
Consequently, the present analysis is intentionally restricted to the \emph{relative} morphology of the diffuse signal rather than its absolute surface brightness. 

As shown in Figure~\ref{original}, the left side of the lunar disk is illuminated by sunlight reflected from the Earth. In contrast, the dark side of the Moon should be essentially black, as there are no significant sources of illumination on that side.
Therefore, the nonzero signal detected on the dark side of the Moon is attributed to bias, dark current, or other instrumental contributions. We thus subtract, for each channel, the mean signal measured over the dark side of the lunar disk from the entire image as a zero-point correction.
Examples of past observations that have used the dark regions of a satellite, where sunlight does not reach, as reference points for determining background radiation include the detection of the X-ray background using the Moon \citep{Schmitt1991} and attempts to detect the extragalactic background light in the near-infrared using eclipses of the Galilean satellites of Jupiter \citep{Tsumura2014}.

\begin{figure*}[ht!]
\plotone{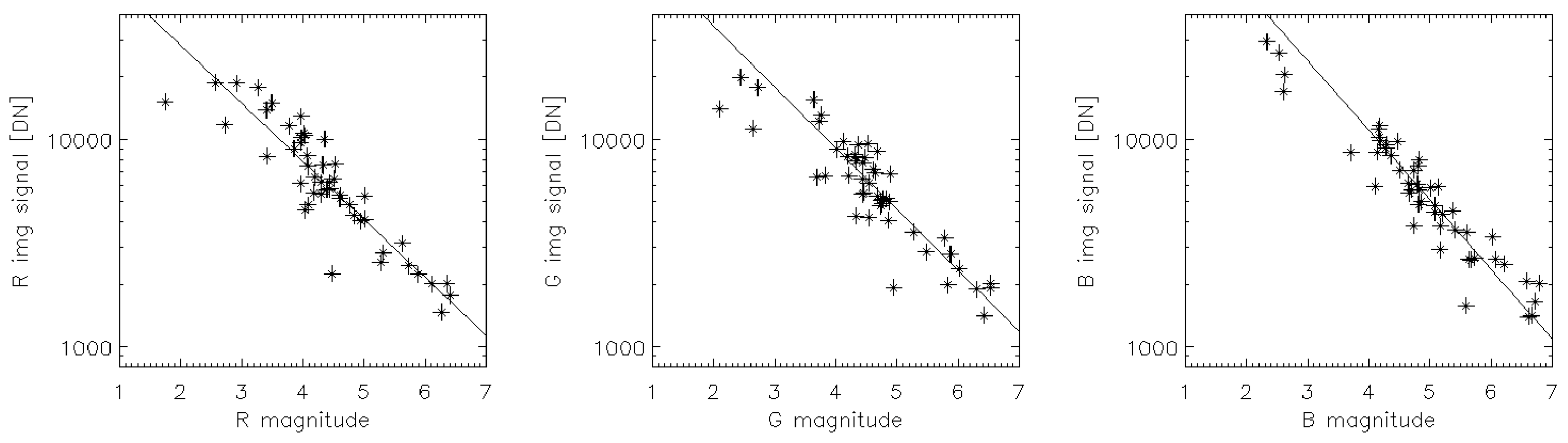}
\caption{
Photometric relation between the cataloged magnitudes of detected stars \citep{cardiel2021_rgb} and their instrumental signals in Data Number (DN) units in the red (left), green (center), and blue (right) channels.
Each solid line represents the result of the best linear approximation in regions darker than 4~mag.
}
\label{starcal}
\end{figure*}

\subsection{Relative Photometric Calibration Using Stars}
\label{subsec:starcal}

It is important to verify that the image retains a monotonic and usable photometric response in the unsaturated regime.
Therefore, the bright field stars in the image were identified and compared with their instrumental RGB signals, which were measured in data numbers (DN) units, with cataloged RGB magnitudes.
The synthetic RGB photometric standards compiled by \citet{cardiel2021_rgb} were adopted as a practical reference, as a bright-star reference specifically intended for consumer-grade digital cameras. The stellar point-spread functions (PSF) extend over several dozen pixels, with a few central pixels approaching saturation at approximately 4.5, 4.8, and 4.1~mag in the red, green, and blue channels, respectively, owing to the image's shallow 8-bit depth. 
Nevertheless, we confirm that the instrumental response of the total stellar signal remains approximately linear in all three color channels (Figure~\ref{starcal});
most pixels comprising each stellar image remain well below the saturation threshold, and thus, the measured total signals are dominated by the detector's linear response regime.

Let $S$ denote the signal value of a detected star measured from the JPEG image and $m$ represent the stellar magnitude. The empirical linear relation obtained from the stellar measurements (Figure~\ref{starcal}) can be expressed as $\log S = -a m$.
The obtained values of the slopes $a$, derived using stars fainter than 4~mag in each channel, are 0.28 (R), 0.29 (G), and 0.33 (B), respectively.
As the flux $F$ and magnitude are related by $m = -2.5 \log F$, the slope of a linear relation between $\log F$ and $m$ is $-1/2.5 = -0.4$. 
These relations imply that the conversion from RAW images, whose pixel values should be proportional to the received flux up to the saturation point, to the JPEG image involves a nonlinear transformation that can be approximated as $S \propto F^{\Gamma}$, where $\Gamma = a/0.4$. 
Such a power-law relation is consistent with the tone-curve (gamma) correction typically applied during RAW-JPEG conversion in consumer digital cameras. 
The derived $\Gamma$ are shallower than the nominal standard RGB $\Gamma$ (1/2.2 = 0.45), which is expected as consumer digital cameras apply camera-specific tone curves during RAW-JPEG conversion.
As a result, the effective encoding $\Gamma$ typically deviates from an idealized standard RGB transformation and can vary between color channels. 
Regardless of the specific transformations applied during the JPEG image creation process, we empirically established the correlation between the JPEG signal values and the actual stellar brightness using field stars. We therefore used this relation to convert the JPEG images into images representing relative sky brightness.

The diffuse signals analyzed in this study are $<$200~DN even at their brightest points, which is sufficiently fainter than the saturation threshold.

\subsection{Masking and Normalization}
\label{subsec:masking}

\begin{figure*}[ht!]
\plotone{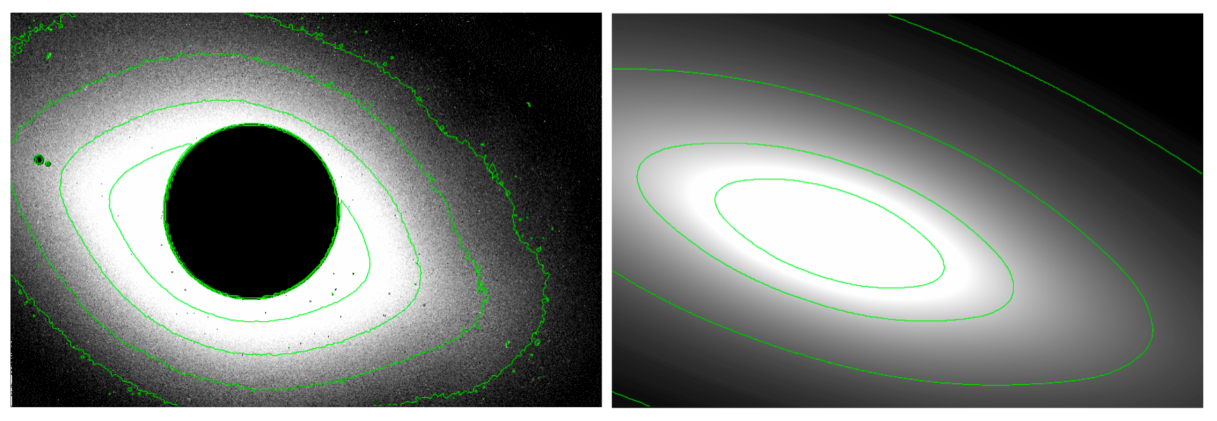}
\caption{
Left: A normalized green channel image produced by masking the Moon and bright stars from the original \texttt{art002e009301} image to extract the structure of the diffuse background emission.
Right: A diffuse ZL map for the same field calculated using \texttt{ZodiSURF} \citep{OBrien_2026} with the same normalization.
Intensity contours of the diffuse light at levels of [0.01, 0.03, 0.1, 0.3] are overlaid for both images.
}
\label{contour}
\end{figure*}

Our primary interest is the structure of the F-corona behind the Moon.
Therefore, in each image, we masked the lunar disk, bright field stars, and other bright features such as stray light or hot pixels. 
The masked images in each color channel were then normalized, as the present analysis is intentionally restricted to the relative morphology of the F-corona owing to the absence of bias and dark-current subtraction. 
Figure~\ref{contour} (left) shows the masked and normalized green channel image with intensity contours overlaid.
Similar contour maps were obtained for the red and blue channels.

\section{Results and Discussion} \label{sec:result}

\subsection{Overall F-corona Shape} \label{sec:color}

The intensity distribution of the F-corona exhibits a flattened, elliptical morphology aligned with the ecliptic plane (Figure~\ref{contour} [left]). 
To quantify the degree of oblateness, we used the commonly adopted flattening index $f$ \citep{Koutchmy1985, Stenborg2018}:
\begin{equation}
f = \frac{R_{\rm eq}}{R_{\rm pol}} - 1,
\end{equation}
where $R_{\rm eq}$ and $R_{\rm pol}$ are the lengths of the semimajor and semiminor axes, respectively. 
The flattening indices were measured to be 0.52, 0.54, and 0.56 from the intensity contours of the red, green, and blue channels, respectively. 
In addition, there is a noticeable asymmetry in the intensity distribution; i.e., the northern and western sides are slightly brighter than the southern and eastern sides.

\begin{figure}[hbt!]
\plotone{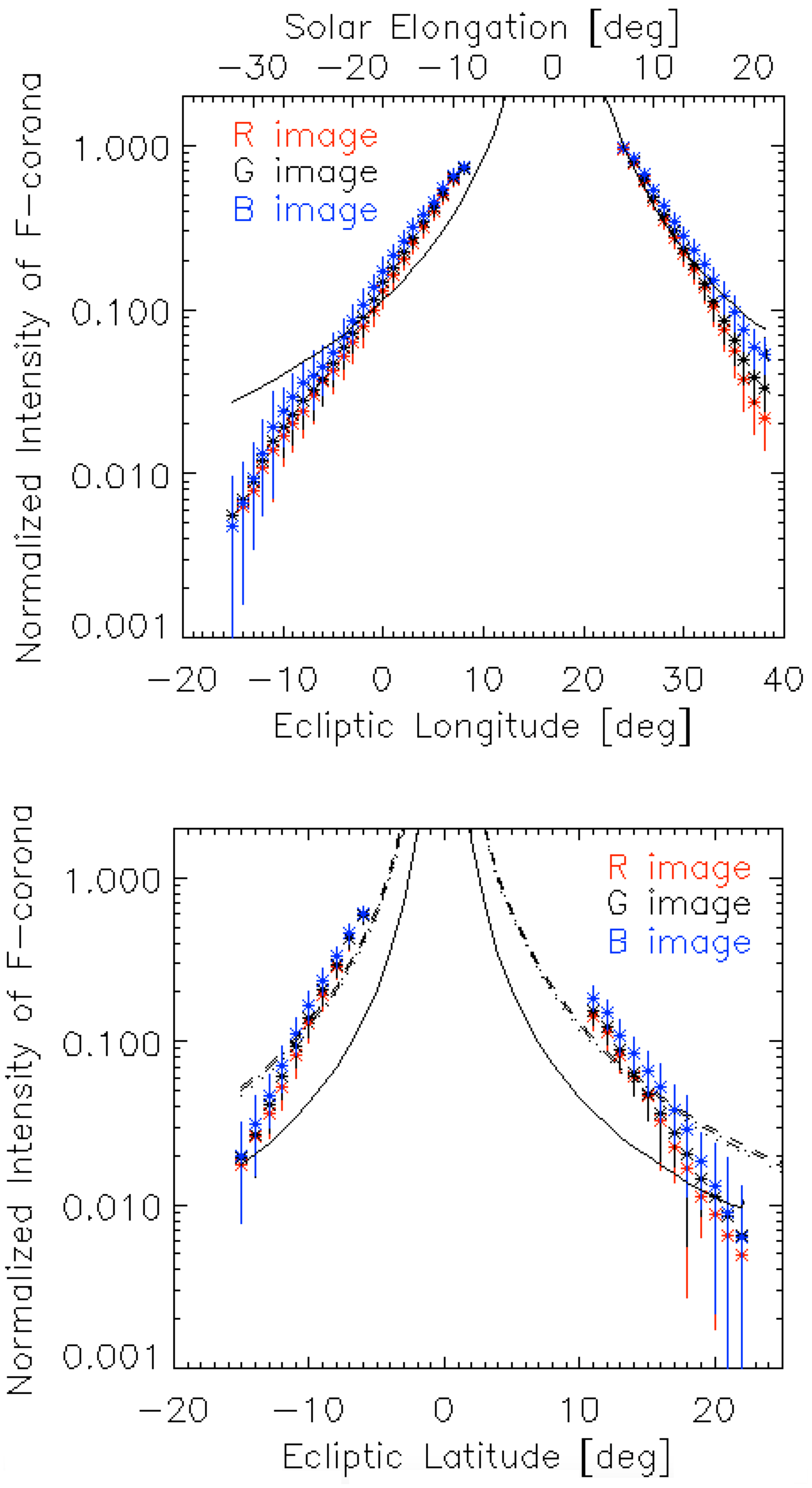}
\caption{
Radial intensity profiles of the F-corona measured from the solar position along the ecliptic longitude (top) and ecliptic latitude (bottom) directions.
An additional axis showing the solar elongation has been included in the top panel, while in the bottom panel the elongation corresponds to the absolute value of the ecliptic latitude.
Red, black, and blue symbols denote the profiles in red, green, and blue channels, respectively.
Error bars represent the standard deviation of pixel values within each spatial bin.
The solid curve shows the ZL profiles calculated with \texttt{ZodiSURF} \citep{OBrien_2026}.
The dotted, dashed, and dash-dot curves in the bottom panel
represent three cases in which two of the three parameters ($\eta$, $\gamma$, and $\mu$) are fixed at the original values in \citet{Kelsall1998}, while the other is varied with $\eta=2.0$, $\gamma=3.0$, and $\mu=1.0$, respectively.
}
\label{fig:profile}
\end{figure}

Figure~\ref{fig:profile} presents the radial intensity profiles of the F-corona in each color channel, which was measured along the ecliptic longitude (east--west) and latitude (north--south) directions from the solar position. 
For the profile along ecliptic longitude, we computed the mean and standard deviation of pixel values within a $1^\circ$-wide longitude bin and $\pm 0.5^\circ$ ecliptic latitude at each elongation. 
These bins were then stepped outward from the solar position in $1^\circ$ increments. 
The profile along the ecliptic latitude was constructed analogously.
The error bars represent the standard deviation of pixel values within each spatial bin and reflect the statistical scattering of the diffuse emission; systematic uncertainties related to the spectral response of the camera were not included, but such effects would have a minimal impact on the relative profile shapes discussed here.

The radial intensity profiles of the F-corona are broadly similar across the RGB channels.
The radial profiles were fitted with power-law functions of solar elongation for each color channel and direction. 
To ensure consistency, the fitting was performed over a solar-elongation range from $10^\circ$ to $15^\circ$, where data are available in all four directions.
The resulting power-law exponents are summarized in Table~\ref{tab:exponents}. 
The eastern and southern profiles exhibit steeper gradients than the western and northern profiles, respectively.
In particular, a significant north--south asymmetry in the F-corona brightness was observed.
This observed north--south asymmetry can be naturally interpreted in the context of the viewing geometry of the interplanetary dust cloud \citep{Lamy2022}. 
Because the observation is made from a location that is not exactly aligned with the symmetry plane of the zodiacal dust distribution,
such an offset can naturally produce an apparent north--south asymmetry in the observed brightness, depending on the observer’s position and line of sight.
In particular, near early April, the Earth is located close to its maximum northern displacement relative to the symmetry plane of the dust cloud, leading to enhanced brightness in the northern direction. 
Accordingly, the directions defined in terms of ecliptic longitude and latitude in Figure~\ref{fig:profile} do not coincide with the intrinsic symmetry axis of the zodiacal cloud, which contribute to the observed directional differences in the radial profiles.

\begin{deluxetable}{ccccc}
\tablecaption{Power-law exponents of the radial intensity profiles within a range of solar elongation from $10^\circ$ to $15^\circ$
\label{tab:exponents}}
\tablehead{
\colhead{Color} & \colhead{East} & \colhead{West} & \colhead{North} & \colhead{South}
}
\startdata
R & -2.88 & -2.84 & -3.69  & -4.73 \\
G & -2.82 & -2.68 & -3.85  & -4.88 \\
B & -2.48 & -2.37 & -3.35  & -5.05 
\enddata
\end{deluxetable}

\subsection{Comparison with Previous Observations} \label{sec:past}

The Artemis~II image is mostly consistent with previous space-based observations of the F-corona. 
The diffuse emission exhibits a flattened, nearly elliptical morphology that is approximately aligned with the ecliptic plane, closely resembling the Moon-occulted view obtained by Clementine \citep{Hahn2002}
and the reconstructions from STEREO-A/HI-1A \citep{Stenborg2018}. 
The flattening indices derived in this study ($f$ = 0.52-0.56) fall within the reported range from STEREO-A/HI-1A ($f \sim$ 0.46-0.66) \citep{Stenborg2018}, indicating that the large-scale geometry seen in the Artemis~II image is representative of the inner zodiacal cloud. 
Within the precision permitted by the present normalized contour analysis, we did not find evidence for strong localized asymmetries or ring-like enhancements, which is in agreement with the morphology established by LASCO observations \citep{Lamy2022}.

STEREO-A/HI-1A observations showed that the radial intensity profiles follow power laws in solar elongation, with exponents ranging from $-2.31$ to $-2.35$ along the east--west direction and from $-2.45$ to $-2.53$ along the north--south direction, respectively \citep{Stenborg2018}. 
Furthermore, LASCO and WISPR demonstrated that these behaviors connect smoothly from the classical F-corona to the outer ZL \citep{Lamy2022, Stenborg2021}. 
Our results are systematically steeper than the STEREO-A/HI-1A values, corresponding to larger absolute values of the exponent, particularly in the north--south direction.

We do not attempt a rigorous comparison of absolute surface brightness levels since the present analysis is based on a rendered RGB JPEG image and is intentionally restricted to relative photometry.
However, the Artemis~II profiles reproduce the expected stronger emission concentration toward the ecliptic plane relative to higher ecliptic latitudes, providing an independent wide-field confirmation of the canonical flattened morphology of the F-corona.

\subsection{Assessing Possible K-corona Contamination} \label{K-corona}

In addition to the F-corona, the solar corona contains a component produced by
Thomson scattering of photospheric light by free electrons, known as the K-corona.
Separating the F-corona from the K-corona is an essential consideration in coronal observations.
The K-corona is linearly polarized by Thomson scattering, whereas the F-corona is nearly unpolarized owing to forward scattering by IDPs;
therefore, the two components have traditionally been distinguished using polarimetric observations \citep{Hulst1950,Lamy2020,Hanaoka2021}. 
Such observations have consistently shown that the brightness of the K-corona decreases faster with solar elongation than that of the F-corona;
the two components become comparable at $\sim$4--6$\,R_{\odot}$), beyond which the F-corona dominates \citep{Hulst1950,Hayes2001}.
The F-corona analyzed herein corresponds to regions at solar elongations $> 7^{\circ}$ (equivalent to distances greater than $25\,R_{\odot}$), wherein the contribution from the K-corona is expected to be negligible.

One line of evidence that contamination from the K-corona is negligible is the absence of strong color-dependent structural differences in the observed F-corona. 
Since the K- and F-coronae exhibit distinct spectral characteristics, any significant contamination from the K-corona would be expected to introduce measurable color-dependent variations associated with its structure \citep{Boe2021}.
However, no such variations are observed. 
This result therefore provides independent support for the conclusion that contamination from the K-corona is negligible.

\begin{figure}[tb!]
\plotone{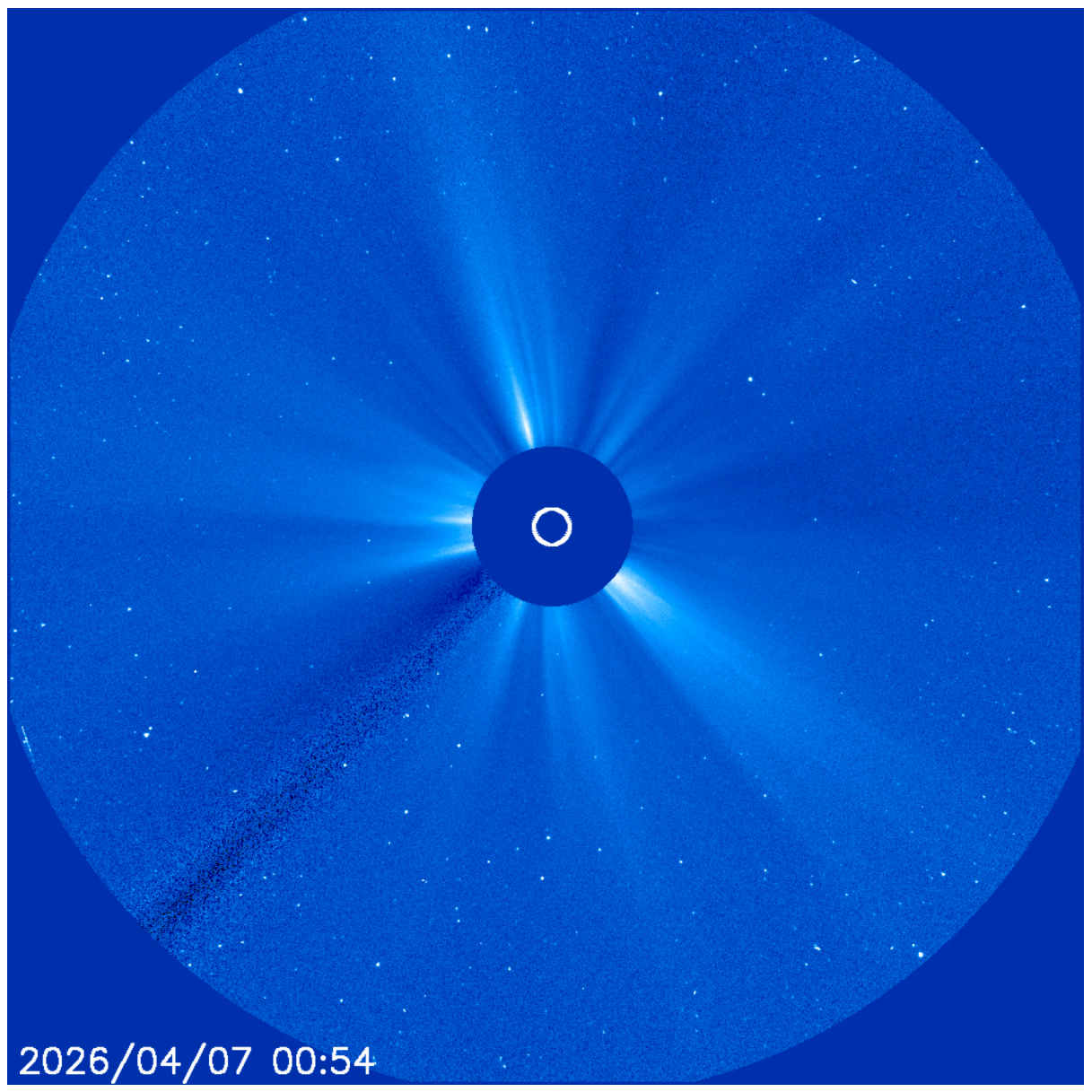}
\caption{
An image obtained by the LASCO~C3 coronagraph aboard SOHO at 00:54~UTC on April~7, 2026, approximately 12 min before the solar-eclipse image analyzed in this study was captured. The FoV of the LASCO~C3 image is $15.7^{\circ}$, comparable to the apparent lunar diameter ($16.9^{\circ}$) as shown in Figure~\ref{fig:streamer}. The central occulter blocks the direct solar disk, with the white circle indicating the position of the Sun. The bright radial structures extending outward from the Sun are coronal streamers.
}
\label{fig:LASCO}
\end{figure}

\begin{figure*}[hbt!]
\plotone{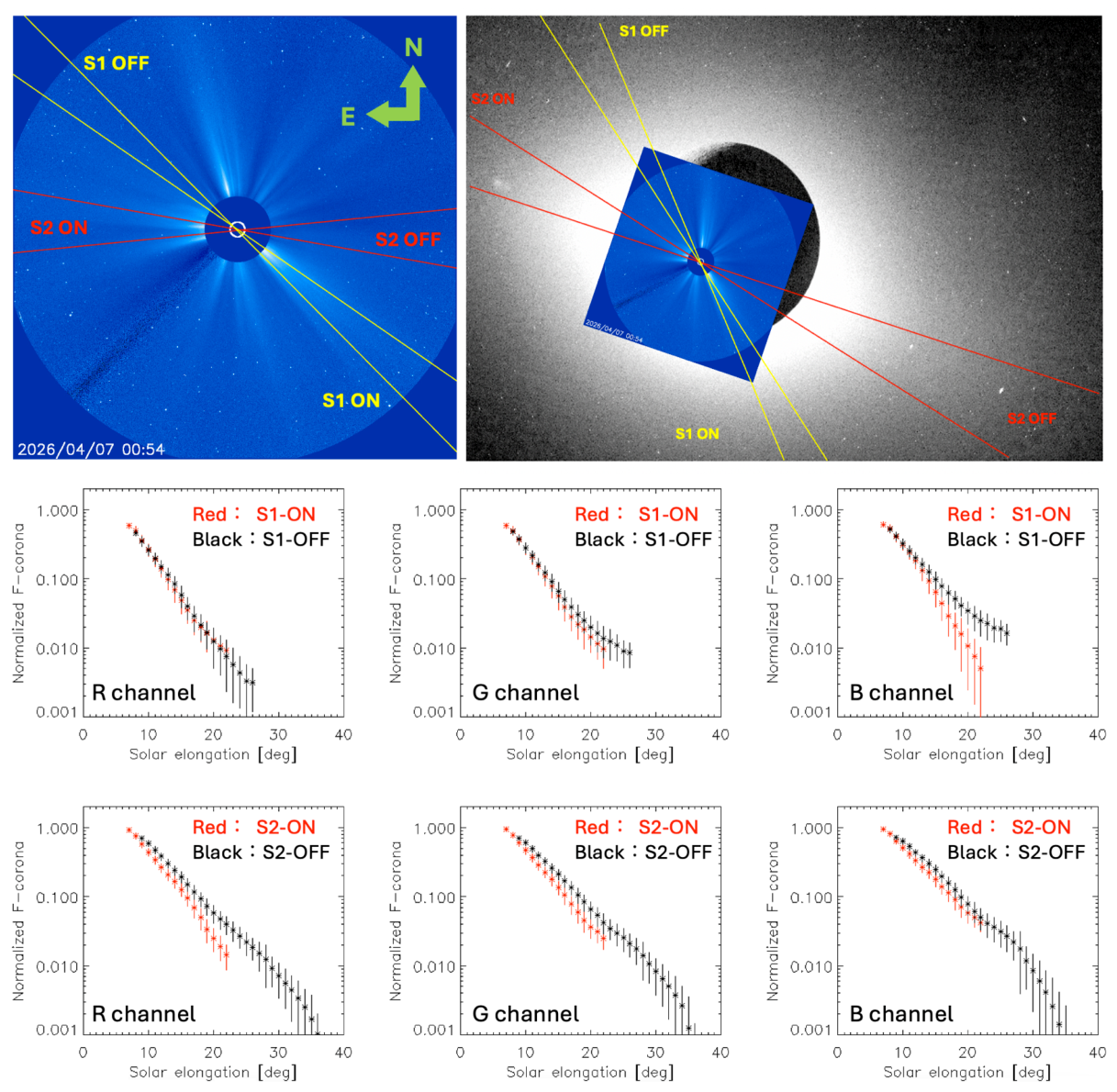}
\caption{
Top left: LASCO~C3 image (same as Figure~\ref{fig:LASCO}) with the prominent coronal streamers, S1 and S2, marked. 
Top right: Solar eclipse image (same as Figure~\ref{original}) with the LASCO image superimposed and aligned in both coordinate system and scale.
Middle and bottom: Radial brightness profiles of the F-corona in each color channel, measured along the streamer and opposite directions. The middle panels correspond to the S1 streamer, whereas the bottom panels correspond to the S2 streamer.
}
\label{fig:streamer}
\end{figure*}

Nevertheless, we performed an additional comparison using a LASCO image to further examine the possibility of contamination from the K-corona. 
Figure~\ref{fig:LASCO} shows an image obtained by the LASCO C3 coronagraph at 00:54~UTC on April~7, approximately 12 minutes before the solar-eclipse image was captured by the Artemis~II crew. The FoV of LASCO C3 spans $15.7^{\circ}$ (i.e., up to $\sim$30$\,R_{\odot}$) \citep{Brueckner1995}.
The bright radial structures emanating from the Sun in the LASCO image are coronal streamers, which trace regions of enhanced electron density and are therefore associated with stronger K-coronal emission. 
To assess a possible residual contribution from the K-corona, we compared the radial brightness profiles of the F-corona measured along the streamer directions with those measured in the opposite directions.
We selected prominent, well-defined streamers extending toward the southwest (S1) and eastward (S2) that had no comparable streamer structures present in the opposite directions, as shown in the top panels of Figure~\ref{fig:streamer}. 
The radial brightness profiles, measured along the streamer direction and along the opposite direction, of the F-corona in each color channel are shown in the middle (S1) and bottom (S2) panels of Figure~\ref{fig:streamer}. 
If there was considerable contamination from the K-corona, the brightness along the streamer directions would be expected to exceed those in the opposite directions. 
However, the brightness measured along the streamer directions was slightly lower than that in the opposite directions across all color channels, indicating that, within the analyzed region, any contribution from the K-corona is negligible. 
The physical origin of the slight dimming of the F-corona observed along the streamer directions remains unclear: 
one possible explanation is that dynamic processes associated with coronal streamers may reduce the local IDP density through destruction or outward transport, although detailed investigation of this effect is beyond the scope of the present study.

\subsection{Comparison with Zodiacal Light Models} \label{sec:zodimodel}

We compared the observed spatial distribution of the F-corona with the ZL distribution calculated using the \texttt{ZodiSURF} model \citep{OBrien_2026}. 
Figure~\ref{contour} (right) shows the ZL map predicted by the model for the same field, whereas Figure~\ref{fig:profile} compare the model radial profiles and observed values. 
The comparisons indicate that the observed F-corona is more extended than predicted by the \texttt{ZodiSURF} model along the ecliptic latitude. 
The \texttt{ZodiSURF} model is calibrated primarily using COBE/DIRBE observations obtained at large solar elongation angles ($>80^{\circ}$) \citep{Kelsall1998};
therefore, it is formally validated only in that regime and is not expected to quantitatively reproduce the structure of the F-corona observed at smaller elongation angles.
Rather than treating this discrepancy as a failure of the underlying functional form, we use the \texttt{ZodiSURF} framework as a phenomenological reference and explore, whether variations in the model parameters can better reproduce the observed brightness morphology of the inner ZL.

In the \texttt{ZodiSURF} framework, the ZL intensity $I_{\rm ZL}$ is calculated as the line-of-sight integral of scattered sunlight,
\begin{equation}\label{eq:ZLmodel}
I_{\rm ZL} = \int F_{\odot}(r)\, n(r,\zeta)\, A\, \Phi(\theta)\, dl,
\end{equation}
where $F_{\odot}(r) \propto r^{-2}$ is the solar flux at heliocentric distance $r$, $n(r, \zeta)$ is the IDP number density at $r$ and $\zeta = |\sin \beta|$ with $\beta$ denoting the ecliptic latitude, $A$ is the particle albedo, $\Phi(\theta)$ is the scattering phase function at scattering angle $\theta$, and $dl$ denotes an infinitesimal path length along the line of sight. 
We then focus on modifying the assumed number-density distribution $n(r, \zeta)$
because variations in $A$ or $\Phi(\theta)$ alone cannot readily reproduce the observed north--south and east--west asymmetries.

The IDP number density is assumed to be separable into radial and vertical components,
\begin{equation}
n(r,\zeta) = n_{0}\left(\frac{r}{r_{0}}\right)^{-\alpha} f(\zeta),
\end{equation}
where $n_{0}$ is the reference number density at heliocentric distance $r_{0}$, $f(\zeta)$ describes the vertical distribution, and $\alpha$ is a free parameter. 
The vertical term $f(\zeta)$ is expressed using a widened, modified fan model,
\begin{equation}
f(\zeta) = \exp(-\eta g^{\gamma}),
\end{equation}
with
\begin{equation}
g =
\begin{cases}
\zeta^{2}/2\mu, & \zeta < \mu, \\
\zeta - \mu/2,  & \zeta \ge \mu,
\end{cases}
\end{equation}
where $\eta$, $\gamma$, and $\mu$ are free parameters \citep{Kelsall1998}. 
The ZL distribution was recomputed by modifying these free parameters to assess whether the resulting model can empirically reproduce the F-corona's observed morphology.

Along the ecliptic longitude, the radial intensity profile is primarily sensitive to the parameter $\alpha$. 
A value of $\alpha = 1$ is theoretically expected under the Poynting-Robertson effect \citep{Burns1979}.
On the other hand, ZL observations in the inner solar system have reported larger values, with $\alpha \approx 1.3$ inferred from Helios~1/2 \citep{Leinert1981} and Hayabusa2\# (the Hayabusa2 Extended Mission, “Sharp”) observations \citep{Tsumura2023}. 
The \texttt{ZodiSURF} model adopts $\alpha = 1.34$ based on the analysis by \citet{Kelsall1998}.
In the horizontal direction, our data generally agree with the model predictions, although the detailed shape of the radial profile is not fully reproduced.
This result supports a radial dust number-density power-law index of $\alpha \approx 1.3$ even in regions near the Sun.

The discrepancy between the observed profiles and nominal model predictions is more pronounced along ecliptic latitude.
In this case, $\eta$, $\gamma$, and $\mu$, must be varied simultaneously to align the model with experimental observations. 
However, these parameters are strongly degenerate within the present dataset, such that we were unable to identify a unique or optimal parameter set. 
Figure~\ref{fig:profile} (bottom) shows the results for three cases in which two of the three parameters are fixed at the original values from \citet{Kelsall1998}, while the remaining parameter is varied, with $\eta = 2.0$, $\gamma = 3.0$, and $\mu = 1.0$.
All of these curves reproduce the observed relative profiles to a similar degree, in contrast to the standard \texttt{ZodiSURF} parameter set [$\eta, \gamma, \mu$] = [4.14, 0.942, 0.189] adopted from \citet{Kelsall1998}.

\section{Summary} \label{sec:summary}

This study analyzed a total solar eclipse image captured via a
consumer camera by Artemis~II astronauts during the lunar flyby and investigated the F-corona's large-scale structure.
By exploiting the natural suppression of stray light provided by the lunar disk, the F-corona brightness distribution was measured in a regime that remains challenging for conventional space-based coronagraphs.

The analyzed image was a publicly released RGB JPEG image published by NASA.
A key aspect of the analysis was the empirical validation of the instrumental response using field stars in the same images (Section~\ref{subsec:starcal}).
After explicitly accounting for the gamma correction inherent to the JPEG data format, the stellar calibration demonstrated that the instrumental response was well described by a linear relation within the unsaturated regime relevant to our measurements. The signal levels of the observed F-corona fall entirely within this validated linear range, justifying the restriction to empirically established sensitivity limits without invoking additional nonlinearity corrections.

The derived brightness profiles and morphology of the F-corona provide an independent, wide-field confirmation of its canonical flattened structure (Section~\ref{sec:color}).
The radial profiles are systematically steeper than those reported in previous observations, and the profiles along the eastern and southern directions exhibit steeper gradients than those along the western and northern directions, respectively (Section~\ref{sec:past}).
Our result supports a radial dust number-density power-law index of $\alpha \approx 1.3$ even in regions near the Sun (Section~\ref{sec:zodimodel}).
The contribution from the K-corona is expected to be negligible (Section~\ref{K-corona}).

Overall, this study demonstrates that opportunistic observations from crewed lunar missions can provide valuable insights into the structure of the inner zodiacal cloud.
Beyond the specific measurements presented here, this study belongs within a long heritage of lunar-based coronal studies, from early Surveyor observations \citep{Norton1967, Bohlin1971} through Clementine lunar occultations \citep{Hahn2002}, and directly connects them to recent proposals advocating routine coronal observations from lunar orbit \citep{Habbal2013, Cooper2023}.
These analysis provide an empirical demonstration of the scientific viability of such lunar occultation concepts and can be regarded as a valuable proof-of-concept supporting future lunar-orbit coronal missions.

\begin{acknowledgments}
We thank the Artemis~II mission astronauts for capturing this unique image, as well as all members of the Artemis~II mission team who enabled its immediate public release.
The SOHO/LASCO data used here are produced by a consortium of the Naval Research Laboratory (USA), Max-Planck-Institut fuer Aeronomie (Germany), Laboratoire d'Astronomie (France), and the University of Birmingham (UK). SOHO is a project of international cooperation between ESA and NASA.
The authors thank James P. Zibin for providing valuable information and engaging in helpful discussions regarding gamma correction in JPEG images, and Richard G. Arendt for providing valuable information on the zero-point correction.
\end{acknowledgments}

\begin{contribution}
K.T.\ performed the data analysis and led the writing and submission of the manuscript.
K.A.\ contributed to the data analysis and manuscript preparation.
All authors discussed the results and approved the final manuscript.
\end{contribution}

\facility{Artemis II, SOHO}

\software{
\texttt{ZodiSURF} \citep{OBrien_2026},
\texttt{Astrometry.net} \citep{Lang2010},
\texttt{IDLAstro} \citep{Landsman1995},
\texttt{JPL Horizons} \citep{Giorgini1996}
}

\bibliography{Artemis_ApJL_Tsumura}{}
\bibliographystyle{aasjournalv7}

\end{document}